\documentclass[twocolumn]{aa}
\usepackage{graphicx}
\usepackage{txfonts}
\begin{document}
\title{The Minimum Mass for Opacity-Limited Fragmentation in Turbulent Cloud Cores}

\author{ D.F.A.Boyd \inst{1} \and A.P.Whitworth \inst{2} }

\offprints{D.F.A.Boyd, \\
\email{D.Boyd@astro.cf.ac.uk} \\
\email{A.Whitworth@astro.cf.ac.uk}}

\institute{Department of Physics \& Astronomy, Cardiff University, PO Box 913, 
5 The Parade, Cardiff CF24 3YB, Wales, UK}

\date{Received ??, 2004; accepted ??, 2004}

\abstract{
We present a new analysis of the minimum mass for star formation, based on 
opacity-limited fragmentation. Our analysis differs from the standard one, 
which considers hierarchical fragmentation of a 3-D medium, and yields $M_{_{\rm MIN}} 
\sim 0.007\;{\rm to}\;0.010\,M_\odot$ for Population I star formation. Instead 
we analyse the more realistic situation in which there is one-shot fragmentation 
of a shock-compressed layer, of the sort which arises in turbulent star-forming 
clouds. In this situation, $M_{_{\rm MIN}}$ can be smaller than $0.003 M_\odot$. 
Our analysis is more stringent than the standard one in that (a) it requires 
fragments to have condensation timescales shorter than all competing mass 
scales, and (b) it takes into acount that a fragment grows by accretion 
whilst it is condensing out, and has to radiate away the energy dissipated 
in the associated accretion shock. It also accords with the recent detection, 
in young star clusters, of free-floating star-like objects having masses as 
low as $0.003\,{\rm M}_\odot$.

 }

\maketitle


\section{Introduction}

3-dimensional opacity-limited hierarchical fragmentation yields a minimum stellar 
mass $M_{_{\rm MIN3}} \sim 0.007\;{\rm to}\;0.010\,M_\odot$. This result was first 
obtained to order of magnitude by Rees (1976), but the range quoted above is the 
result of more detailed treatments of the thermodynamics of fragmentation by Low 
\& Lynden-Bell (1976), Silk (1977) and Boss (1988). However, there are problems 
with the process of 3-dimensional hierarchical fragmentation, and this estimate 
of $M_{_{\rm MIN3}}$ almost certainly needs to be revised upwards (see 
\S\ref{sec:3D}). 
Moreover, it is now generally accepted (e.g. Elmegreen 2000, Pringle et al. 
2001, Hartmann et al. 2001, Elmegreen 2002, Padoan \& Nordlund 2002, 
V\'azquez-Semadeni et al. 
2003) that star-forming clouds are transient turbulent entities, and that star 
formation occurs almost as soon as a cloud forms, within a few crossing times. In 
this picture, prestellar cores are created where turbulent elements collide with 
sufficient ram-pressure for the resulting shock-compressed layer to be 
gravitationally unstable. The process of fragmentation is then quite 
different from that envisaged in 3-dimensional hierarchical fragmentation. In 
particular, (i) the fragmentation of a shock-compressed layer can, and usually 
does, proceed whilst the layer is still accumulating (and therefore it is 
confined by ram pressure); (ii) the convergent motions leading to the formation 
of a fragment are initially concentrated in the plane of the shocked layer, i.e. 
fragmentation of a shock-compressed layer is essentially two-dimensional.

In this paper, we present a model to describe the 2-dimensional opacity-limited 
fragmentation of a shock-compressed layer and use this model to determine the minimum 
mass of the fragments that can condense out of such a layer, $M_{_{\rm MIN2}}$. 
Section~\ref{sec:3D} reviews the standard paradigm of 3-dimensional hierarchical 
fragmentation, in which a cloud fragments into smaller and smaller subclouds until 
radiative cooling can no longer match the compressional heating rate and the the 
subclouds become approximately adiabatic. In Section~\ref{sec:2D} we highlight the 
differences between hierarchical 3-dimensional fragmentation and the fragmentation 
of a 2-dimensional 
layer. We derive equations to describe how a shock-compressed layer is formed by the 
collision of two streams of gas, and how the layer fragments whilst it is still forming. 
In Section~\ref{sec:results} we use this model to follow the evolution of a fragment and 
hence to estimate the minimum mass for a fragment. In Section~\ref{sec:conclusions} we 
briefly summarize our results.

Observational and numerical evidence for the formation and 
fragmentation of shock-compressed layers, in regions of imminent 
or ongoing star formation, is hard to identify. This contrasts 
with the substantial evidence for the formation and fragmentation 
of filaments in such regions. Nonetheless, we believe that the 
formation and fragmentation of shock-compressed layers must be 
the fundamental process in turbulent star formation regions. First 
(from a theoretical perspective), to create a filament ab initio 
would require a very contrived flow geometry. Second (again from a 
theoretical perspective), the fragmentation of a layer normally 
proceeds via the formation of a network of filaments (e.g. 
Turner et al. 1995, Whitworth et al. 1995, Bhattal et al. 1998); this 
is because of the well established theorem (e.g. Zeldovich 1979) which 
asserts that if gravitational contraction gets ahead in one dimension 
(due either to a greater initial perturbation, or less resistance), 
it tends to get further ahead. Third (from an observational 
perspective), layers are much harder to identify than filaments, 
unless they are seen edge-on, in which case they look like filaments.

Moreover, free-floating objects with masses as low as 
$0.003\,{\rm M}_\odot$ have now been detected in young star 
clusters, like $\sigma$ Orionis (Zapatero Osorio et al. 2002), 
and it beholds us to ask how they might have formed. They cannot 
form by three-dimensional fragmentation, but it appears that 
they could form by layer fragmentation, as analyzed here. Other 
possible formation mechanisms include the photo-erosion of 
pre-existing massive cores (Hester 1997, Whitworth \& Zinnecker 
2004) and impulsive interactions between discs in dense clusters 
(Boffin et al. 1998, Watkins et al. 1998a,b).

\section{Three-dimensional hierarchical fragmentation}\label{sec:3D}

Theories of star formation have traditionally been based on the idea of 
opacity-limited hierarchical fragmentation (Hoyle 1953). A massive, 
3-dimensional, Jeans-unstable cloud starts to contract and, as long as 
the isothermal sound speed, $a$, remains constant, the Jeans mass 
\begin{equation}
M_{_{\rm JEANS3}} \;\simeq\; \left[ \frac{375\,a^6}{4\,\pi\,G^3\,\rho} \right]^{1/2} 
\sim \frac{6\,a^3}{G^{3/2}\,\rho^{1/2}} \,,
\end{equation}
decreases with increasing density, $\rho$. Thus, once the original cloud has 
contracted sufficiently, the Jeans mass $M_{_{\rm JEANS3}}$ is reduced and the cloud 
can fragment into subclouds. Once these subclouds have contracted sufficiently, 
they themselves can fragment into still smaller `sub-subclouds', and so on. 
The process can continue recursively, breaking the original cloud up into 
ever smaller fragments, as long as the isothermal sound speed remains 
approximately constant, i.e. as long as the contracting gas is able to radiate 
away, immediately, the $PdV$ work being done on it by compression. Once 
the smallest fragments become so dense and opaque to their own cooling 
radiation that they can no longer radiate away the $PdV$ work immediately, 
the sound speed starts to increase, the Jeans mass decreases no further, and 
fragmentation therefore stops.

The minimum mass for star formation can be found from a general analysis of 
this process, as shown by Rees (1976); for simplicity we neglect all purely 
numerical factors. A fragment of radius $R$, which has just become Jeans unstable 
and started to condense out, contracts at speed $dR/dt \sim -\,a$. Hence the heating 
rate due to $PdV$ work is
\begin{equation}
{\cal H} \;=\; -\,P\,\frac{dV}{dt} \;\sim\; -\,\rho\,a^2\,R^2\,\frac{dR}{dt} 
\;\sim\; \rho\,a^3\,R^2 \,,
\end{equation}
where we have substituted $P = \rho a^2$ and  $dV/dt \sim R^2 dR/dt$. If the 
fragment is to remain approximately isothermal, this must be less than the 
maximum possible radiative cooling rate,
\begin{equation}
{\cal C} \;\sim\; R^2\,\sigma_{_{\rm SB}}\,T^4\,,
\end{equation}
where $\sigma_{_{\rm SB}}$ is the Stefan Boltzman constant and $T$ is the temperature. 
This maximum cooling rate will only be realized if the fragment cools like a blackbody, 
i.e. if it is optically thick but only just; detailed calculations (e.g. Low \& 
Lynden-Bell, 1976) indicate that this is usually the case. To ensure that 
$\mathcal{C}>\mathcal{H}$, i.e. that cooling is sufficently rapid to keep $a$ 
constant, we require
\begin{equation}
\rho \;\la\; \rho_{_{\rm MAX}} \;\sim\; \frac{\sigma_{_{\rm SB}}\,T^4}{a^3} \,,
\end{equation}
and hence
\begin{equation}
M \;\ga\; M_{_{\rm MIN3}} \;\sim\; \frac{a^{3}}{G^{3/2}\rho_{_{\rm MAX}}^{1/2}} 
\;\sim\; c\left[ \frac{h}{G}\right]^{3/2}\,\left[\frac{k_{_{\rm B}}T}{\bar{m}^{9}}
\right]^\frac{1}{4} \,,
\end{equation}
or equivalently
\begin{equation} \label{eqn:planck}
M \;\ga\; M_{_{\rm MIN3}} \;\sim\; \frac{m_{_{\rm PLANCK}}^3}{\bar{m}^2} \, 
\left[ \frac{a}{c} \right]^{1/2} \,.
\end{equation}
Here we have substituted $a=(k_{_{\rm B}}T/\bar{m})^{1/2}$, where $k_{_{\rm B}}$ 
is Boltzmann's constant and $\bar{m}$ is the mean gas-particle mass; 
$\sigma_{_{\rm SB}} \sim k_{_{\rm B}}^4/c^2h^3$, where $c$ is the speed of light 
and $h$ is Planck's constant; and $m_{_{\rm PLANCK}} = [ch/G]^{1/2}$ is the 
Planck mass \footnote{Equation (\ref{eqn:planck}) shows that $M_{_{\rm MIN3}}$ 
is much smaller than the Chandrasekhar mass by virtue of (i) the mean gas-particle 
mass $\bar{m}$ being much larger than the mean mass per electron ($2 m_{_{\rm PROTON}} 
/ (1+X)$, where $X$ is the fractional abundance of hydrogen by mass); and (ii) 
the sound speed, $a$, being much less than the speed of light, $c$.}.

However, there are problems with the notion of hierarchical fragmentation. 
Observationally, there is no clear evidence for it occuring, and from a 
theoretical viewpoint, it does not seem to work because the time-scale 
on which a subcloud condenses out is always longer than the time-scale on 
which its parent cloud is contracting.

Specifically, the condensation timescale in 3-D is given by
\begin{equation}
t_{_{\rm COND3}} \;\simeq\; t_{_{\rm FF3}}\,\left\{ 1\,-\, \left[ 
\frac{M_{_{\rm JEANS3}}}{M} \right]^{2/3} \right\}^{-1/2} \,,
\end{equation}
where $t_{_{\rm FF3}} = (3\pi/32 G \rho)^{1/2}$ is the freefall time in 3D. 
Thus a parent cloud is always closer to freefall collapse than its 
subclouds. If we assume that the whole of the parent cloud breaks up into 
subclouds, so that at their inception the subclouds are touching, then a 
subcloud has insufficient time to establish itself as a distinct entity before 
it gets merged with neighbouring subclouds by the overall contraction of the 
parent cloud. For example, a subcloud with $M \simeq 4 M_{_{\rm JEANS3}}$ 
condenses out on a timescale $\sim 1.3\,t_{_{\rm FF3}}$ (and smaller subclouds 
condense out even more slowly), whereas the parent cloud contracts on a time-scale 
$\sim t_{_{\rm FF3}}$.

If, instead, we assume that at their inception the subclouds are not touching, 
we must consider whether a subcloud continues to grow by accreting material 
from its surroundings as it condenses out. If we take at face value the Bondi 
formula for the rate of spherically symmetric accretion onto a mass $M$, from 
a background medium having density $\rho$ and isothermal sound speed $a$,
\begin{equation}
\frac{dM}{dt} \;=\; \frac{{\rm e}^{3/2}\,\pi\,G^2\,\rho\,M^2}{a^3}
\end{equation}
(Bondi 1952), we can rewrite this equation in the form
\begin{equation}
\frac{d\ell n(M)}{d[t/t_{_{\rm FF3}}]} \;\simeq\; 42\,\left[ 
\frac{M}{M_{_{\rm JEANS3}}} \right] \,.
\end{equation}
Thus, even though the Bondi scenario is not strictly applicable, the implication 
is that the subcloud mass will increase by a very large factor whilst it is 
condensing out (e.g. a fragment with initial mass $4 M_{_{\rm JEANS3}}$ notionally 
increases its mass by a factor of $42 \times 4 \times 1.3 \simeq 220$). We must 
conclude that the initial fragment mass will be a significant underestimate of the 
final fragment mass.

\section{Fragmentation of a two-dimensional layer}\label{sec:2D}

There is an alternative paradigm to hierarchical fragmentation, which 
overcomes these problems and agrees better with observation. In this 
paradigm (e.g. Larson 1981, Elmegreen 2000, Pringle, Allen \& Lubow 2001, 
Hartmann, Ballesteros-Paredes, \& Bergin 2001, Padoan \& Nordlund 2002, 
Elmegreen 2002, V{\'a}zquez-Semadeni, Ballesteros-Paredes \& Klessen 2003, 
Mac Low \& Klessen 2004), giant molecular clouds are relatively 
short-lived objects which form and dissolve on a dynamical timescale and have 
a turbulent and inhomogeneous internal structure. The substructure within 
a molecular cloud is therefore highly transient, with clumps forming, dispersing, 
and re-forming on a dynamical timescale, without necessarily spawning new stars. 
Only occasionally will a particularly dense, massive and strongly converging shock 
lead to the formation of a prestellar core. In this scenario, both core formation 
and core collapse are triggered by the collision of two turbulent elements of gas 
and the formation of a gravitationally unstable, shock-compressed layer. This is a 
one-step, two-dimensional fragmentation process: `one step' because the process 
does not repeat itself hierarchically; `two-dimensional' because, when the layer 
fragments, the wavelengths of the most rapidly growing fragments are much larger 
than the thickness of the layer (Whitworth {\it et al.} 1994a,b).

\subsection{Linear fragmentation of a static layer in plane-parallel symmetry}

2-dimensional fragmentation of a static layer (e.g. Larson 1985) is fundamentally 
different from 3-dimensional fragmentation. The Jeans mass is given by
\begin{equation}
M_{_{\rm JEANS2}} \;\simeq\; \frac{9\,a^4}{16\,\pi\,G^2\,\Sigma} \;\sim\; 
\frac{0.2\,a^4}{G^2 \Sigma} \,,
\end{equation}
where $\Sigma$ is the surface density of the layer; and the timescale on which 
Jeans unstable fragments condense out of the layer is given by
\begin{equation}
t_{_{\rm COND2}} \;\simeq\; \frac{a}{\pi\,G\,\Sigma} \, \left\{ \left[ \frac
{M_{_{\rm JEANS2}}}{M} \right]^{1/2} \,-\, \left[ \frac{M_{_{\rm JEANS2}}}{M} 
\right] \right\}^{-1/2} \,.
\end{equation}
Since $t_{_{\rm COND2}}$ has a minimum for $M \simeq 4M_{_{\rm JEANS2}}$, 
fragments on this mass scale condense out faster than smaller fragments {\it and} 
faster than larger fragments. Therefore $4M_{_{\rm JEANS2}}$ is a preferred 
mass scale for fragmentation. Fragmentation on this scale is likely to be 
permanent, because the fragments will not be merged by the overall contraction 
of the parent layer on larger scales. Given these differences between 
hierarchical fragmentation of a 3-dimensional cloud and one-step fragmentation 
of a 2-dimensional layer, we are interested in exploring how, and at what mass 
scale, opacity limits the fragmentation of a 2-dimensional layer. Is the 
minimum mass formed at the opacity limit significantly different from that 
estimated for 3-dimensional fragmentation?

\subsection{A shock-compressed layer in plane-parallel symmetry}

\begin{figure*} \label{fig:2Dlayer}
\centering
\includegraphics[angle=0,width=9cm]{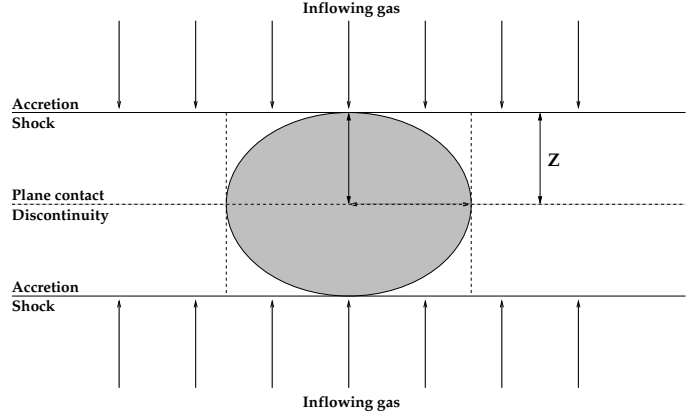}
\caption{Gas streams having density $\rho$ and sound speed $a$ collide to 
create a shock-compressed layer. As gas continues to flow into the layer 
from each side at speed $\pm v$, a spheroidal fragment of initial radius 
$r_{_{\rm INIT}}$ and initial height $z_{_{\rm INIT}} = Z$ begins to 
condense out.}
\end{figure*}

We consider two identical streams of gas with uniform density $\rho$ and uniform 
isothermal sound speed $a$, which collide head-on at relative speed $2v$ to form 
a layer. We assume that a plane contact-discontinuity forms where the two 
streams meet and we fix our main coordinate frame in this contact discontinuity. 
The two streams approach the contact discontinuity with velocities $\pm v$, and 
a plane-parallel layer forms symmetrically about the discontinuity, bounded by 
two accretion shocks, as illustrated in Figure~\ref{fig:2Dlayer}. For simplicity 
we assume that radiative cooling in the shocked gas is so efficient that the gas 
rapidly cools back to its pre-shock temperature. If the full thickness of the 
layer is $2Z(t)$, then the gas flows into the accretion shock at speed 
$u_{_{\rm IN}}=v+\dot{Z}$, where $\dot{Z}=dZ/dt$, and out at speed 
$u_{_{\rm OUT}}=\dot{Z}$; note that $u_{_{\rm IN}}$ and $u_{_{\rm OUT}}$ are 
measured relative to the shock-front. Applying the isothermal shock condition,
\begin{equation}
\frac{u_{_{\rm IN}}}{a} \;=\; \frac{a}{u_{_{\rm OUT}}} \,,
\end{equation}
we obtain
\begin{equation}
\frac{v+\dot{Z}}{a} \;=\; \frac{a}{\dot{Z}} \,, 
\end{equation}
whence
\begin{equation}
\dot{Z} \;=\; \frac{[v^{2}+4a^{2}]^{1/2}-v}{2} \,.
\end{equation}

The post-shock density (i.e. the density in the shock compressed layer) is 
\begin{equation}
\rho_{_{\rm SHOCKED}}\; =\; \rho\,\frac{u_{_{\rm IN}}}{u_{_{\rm OUT}}}=\rho\,
\frac{[v^{2}+4a^{2}]^{1/2}+v}{[v^{2}+4a^{2}]^{1/2}-v} \,,
\end{equation}
and so the surface density $\Sigma$ of the layer grows linearly with time according to 
\begin{equation} \label{eq:sigma}
\Sigma(t) \;=\; 2\,\rho_{_{\rm SHOCKED}}\,\dot{Z}\,t \;=\; \rho\,
\left\{[v^{2}+4a^{2}]^{1/2}+v\right\}\,t \,. 
\end{equation}

As long as $v \gg a$, we can approximate 
\begin{equation}
\dot{Z} \;\simeq\; \frac{a^{2}}{v}\ll a \,,
\end{equation}
\begin{equation}
\rho_{_{\rm SHOCKED}} \;\simeq\; \frac{\rho\,v^{2}}{a^{2}} \,,
\end{equation}
\begin{equation}
\Sigma(t) \;\simeq\; 2\,\rho\,v\,t \,,
\end{equation}
and so the sound-crossing time of the layer, $Z/a$, is much smaller than its 
growth time, $Z/\dot{Z}$. This means that the layer has sufficient time to relax 
and remain close to hydrostatic equilibrium. We also note that, provided
\begin{equation}
t \;<\; \frac{1}{2\,(G\,\rho)^{1/2}} \;,
\end{equation}
$G\Sigma^{2} \ll \rho_{_{\rm SHOCKED}}a^{2}$, and thus self-gravity plays a 
negligible role compared with ram pressure in confining the layer in the 
direction perpendicular to the contact discontinuity. Consequently, the 
plane-parallel hydrostatic equilibrium of the layer has a very flat density 
profile, with little density contrast between the centre and the edge.

\subsection{Linear fragmentation of the layer} \label{sec:linfrag}

Now we consider a small circular patch of radius $r$ on the layer, and consider 
whether it is able to condense out. The radial motion of the patch is determined 
by a competition between the hydrostatic acceleration $\sim a^{2}/r$ (which 
promotes expansion), and the self-gravitational acceleration $\sim \pi G\Sigma$ 
(which promotes contraction). Therefore we can write
\begin{equation}
\ddot{r} \;\simeq\; \frac{a^{2}}{r}-\pi G\Sigma,
\end{equation}
and contraction ($\ddot{r} < 0$) requires
\begin{equation}
r \;>\; R_{_{\rm JEANS2}} \;\simeq\; \frac{a^{2}}{\pi G\Sigma}.
\end{equation}
The time-scale for condensation of a Jeans-unstable fragment is given by
\begin{equation}
t_{_{\rm COND2}} \simeq \left[\frac{r}{-\ddot{r}}\right]^{1/2} \simeq 
\frac{a}{\pi\,G\,\Sigma}\,\left\{ \left[ \frac{R_{_{\rm JEANS2}}}{r} \right] 
\,-\, \left[ \frac{R_{_{\rm \rm JEANS2}}}{r} \right]^2 \right\}^{-1/2}\,,
\end{equation}
so the fastest condensing fragment has initial radius and condensation time-scale
\begin{equation}
r_{_{\rm FASTEST}} \;\simeq\; 2\,R_{_{\rm JEANS2}} \;\simeq\; \frac{2a^{2}}{\pi\,G\,\Sigma},
\end{equation}
\begin{equation}
t_{_{\rm FASTEST}} \;\simeq\; \frac{2a}{\pi\,G\,\Sigma}. \label{eq:tfastest}
\end{equation}
As the layer piles up, $\Sigma$ increases, and so the size of the fastest 
condensing fragment decreases, and its condensation time-scale decreases. 
Condensation into the non-linear regime starts only when $t_{_{\rm FASTEST}} \la t$, 
and substituting for $\Sigma$ in equation (\ref{eq:tfastest}), from equation 
(\ref{eq:sigma}), this condition gives the fragmentation time,
\begin{equation}
t_{_{\rm FRAG}} \;=\; \left\{ \frac{2a}{\pi\,G\,\rho\,\left\{ [v^2+4a^2]^{1/2}+v \right\}} 
\right\}^{1/2} \,.
\end{equation}
The mean radius and mass of the fragments which condense out are therefore
\begin{equation}
r_{_{\rm FRAG}} \;\equiv\; r_{_{\rm FASTEST}}(t_{_{\rm FRAG}}) \;=\; 
\left\{ \frac{2\,a^{3}}{\pi\,G\,\rho \left[ \left(v^2+4a^2\right)^{1/2}
+v \right]} \right\}^{1/2}\,,
\end{equation}
\begin{equation}
m_{_{\rm FRAG}} \;=\; \pi\,r_{_{\rm FRAG}}^2\,\Sigma(t_{_{\rm FRAG}}) \;=\; 
\left\{ \frac{2^3a^7}{\pi\,G^3\,\rho\,\left[ \left( v^2+4^2 \right)^{1/2}+v 
\right]}\right\}^{1/2}.
\end{equation}
We note that
\begin{equation}
\frac{2r_{_{\rm FRAG}}}{2Z(t_{_{\rm FRAG}})} \;\simeq\; \frac{2a}{\left[ 
\left( v^2+4a^2 \right)^{1/2} - v \right]} \;\sim\; \frac{v}{a},
\end{equation}
where the final expression obtains in the limit $v \gg a$. Therefore, at 
its inception, the mean diameter of a fragment is much larger than the 
thickness of the shock-compressed layer.

\subsection{Non-Linear fragmentation of the layer}

The linear analysis above does not take into account the material which continues 
to flow into the shock layer, after it has started to fragment. It also does not 
describe how the fragment develops as it condenses out.

In order to follow the 
condensation of fragments into the non-linear regime, we model them as uniform 
density spheroids with radius $r$ and half-height $z$. The excursions of an 
{\it isolated} oblate spheroid of mass $m$, subjected to external pressure 
$P_{_{\rm EXT}}$, 
are governed by the equations
\begin{eqnarray} \nonumber
\ddot{r} & \simeq & -\, \frac{3\,G\,m}{2} \, \left\{ \frac{r\cos^{-1}(z/r)}
{(r^2-z^2)^{3/2}} \,-\, \frac{(z/r)}{(r^2-z^2)} \right\} \\ \label{eqn:rddot}
 & & \hspace{3.9cm} -\, \frac{20\,\pi\,P_{_{\rm EXT}}\,r\,z}{3\,m} \,+\, 
\frac{5\,a^2}{r} \,, \\ \nonumber
 & & \\ \nonumber
\ddot{z} & \simeq & -\, 3\,G\,m \,\left\{ \frac{1}{(r^2-z^2)} \,-\, 
\frac{z\cos^{-1}(z/r)}{(r^2-z^2)^{3/2}} \right\} \\ \label{eqn:zddot}
 & & \hspace{3.95cm} -\, \frac{20\,\pi\,P_{_{\rm EXT}}\,r^2}{3\,m} \,+\, \frac{5\,a^2}{z} \,,
\end{eqnarray}
In equations (\ref{eqn:rddot}) and (\ref{eqn:zddot}), the first term on the righthand 
side represents self-gravity, the second term represents external pressure, and the 
third term represents internal pressure. Similar expressions describe the excursions 
of a prolate spheroid.

For an oblate spheroidal fragment condensing out of a shock-compressed layer, while 
the layer continues to accrete matter from the converging flows which are forming it, 
we must modify these equations to take into account two facts. (i) The external 
pressure is the ram pressure of the inflowing gas. (ii) The spheroid grows 
in mass as a consequence of the inflowing gas. Knowing $\dot{z}$, which is the speed 
with which the gas already in the spheroid expands, we can calculate how fast, $\dot{y}$, 
the shock front has to advance towards the inflowing gas, in order to decelerate the 
new inflowing gas. For the velocities relative to the shock, we have $u_{_{\rm IN}}=
v+\dot{y}$ and $u_{_{\rm OUT}}=\dot{y}-\dot{z}$, so the isothermal shock condition 
gives
\begin{equation}
\frac{v+\dot{y}}{a} \;=\; \frac{a}{\dot{y}-\dot{z}} \,,
\end{equation}
\begin{equation}
\dot{y} \;=\; \frac{ \left[ \left (v-\dot{z} \right)^2\,+\,4a^2 \right]^{1/2}-
(v-\dot{z})}{2}.
\end{equation}
The external pressure acting on the spheroid in the $z$-direction is therefore 
\begin{eqnarray} \nonumber
P_{_{{\rm EXT},\,z}} & = & \rho\,a^2\,\frac{v+\dot{y}}{\dot{y}-\dot{z}} \\ \label{eqn:Pextz}
 & = & \rho\,a^2\,\frac{ \left[ \left(v-\dot{z} \right)^2 \,+\, 4a^2 \right]^{1/2} \,+\, 
\left( v+\dot{z} \right)}{ \left[ \left (v-\dot{z} \right)^2 \,+\, 4a^2 \right]^{1/2} \,-\, 
\left( v+\dot{z} \right)} \,.
\end{eqnarray}
The external pressure acting on the spheroid in the $r$-direction is the same 
as in the unperturbed layer,
\begin{equation} \label{eqn:Pextr}
P_{_{{\rm EXT},\,r}} \;=\; \rho\,a^2\, \frac{ \left[ v^2+4a^2 \right]^{1/2}+v}
{ \left[ v^2+4a^2 \right]^{1/2}-v}.
\end{equation}

The rate of increase of the mass in the fragment, due to accretion from the 
continuing inflow,  is 
\begin{eqnarray} \nonumber
\dot{m} & = & 2\,\pi\,r^2\,\rho\,\left( v+\dot{y} \right) \\ \label{eqn:mdot}
 & = & \pi\,r^2\,\rho\, \left\{ \left[ ( v-\dot{z})^2+4a^2 \right]^{1/2} \,+\, 
(v+\dot{z}) \right\}\,,
\end{eqnarray}
and we must add a term to equation (\ref{eqn:rddot}) to represent the inertial drag 
of this accreted material, i.e.
\begin{eqnarray} \nonumber
\ddot{r} & \simeq & -\, \frac{3\,G\,m}{2} \, \left\{ \frac{r\cos^{-1}(z/r)}
{(r^2-z^2)^{3/2}} \,-\, \frac{(z/r)}{(r^2-z^2)} \right\} \\ \label{eqn:rddotnew}
 & & \hspace{2.5cm} -\, \frac{20\,\pi\,P_{_{\rm EXT}}\,r\,z}{3\,m} \,+\, 
\frac{5\,a^2}{r} \,-\, \frac{\dot{m}\,\dot{r}}{m} \,.
\end{eqnarray}

The $PdV$ heating rate for the material already in the fragment is 
\begin{equation} \label{eqn:heat}
\mathcal{H} \;=\; -\,P\,\frac{dV}{dt} \;=\; -\,m\,a^2\,\left[ \frac{2\dot{r}}{r}
\,+\, \frac{\dot{z}}{z} \right].
\end{equation}

The heating rate due to dissipation of the $z$-kinetic energy of the matter accreting 
onto the fragment from the inflow, less the $r$-kinetic energy used to accelerate the 
accreted matter laterally, is
\begin{equation} \label{eqn:diss}
\mathcal{D} \;=\; \frac{\dot{m}\,(v+\dot{z})^2}{2} \,-\, \frac{\dot{m}\,\dot{r}^2}{5} \,.
\end{equation}

The maximum (i.e. blackbody) cooling rate from the two sides of the fragment is
\begin{equation} \label{eqn:cool}
\mathcal{C}(T) \;=\; 2\,\pi\,r^2\,\sigma_{_{\rm SB}}\,T^{4} \;=\; \frac{2^2\,\pi^6\,
\bar{m}^4\,a^8\,r^2}{15\,c^2\,h^3},
\end{equation}
where $h$ is Planck's constant and $\bar{m}$ is the mean gas-particle mass, as 
before.

\begin{figure*}
\label{fig:minvrho}
\centering
\includegraphics[angle=-90,width=15cm]{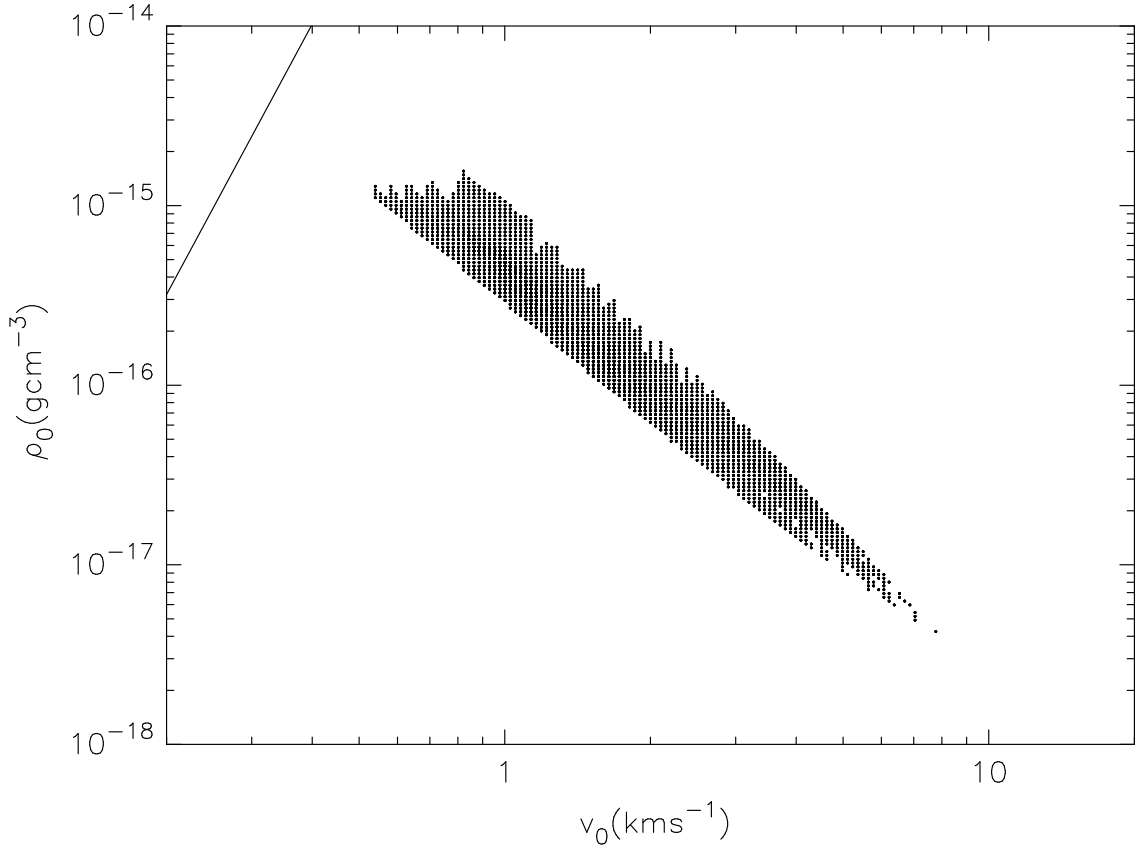} 
\caption{A log/log plot of the ($\rho,v$) plane. The dots mark combinations of 
pre-shock density, $\rho$, and collision speed, $v$, for which (assuming a 
post-shock effective sound speed $a = 0.2\,{\rm km}\,{\rm s}^{-1}$, 
corresponding to molecular gas at $10\,{\rm K}$) the fastest-growing fragment 
mass is less than $0.005 M_\odot$. The irregularities in the boundary of this 
region have to do with the tendency of fragments with mass near the fastest 
growing one to undergo pulsations -- see \S\ref{sec:oscillations} and 
\S\ref{sec:optimum}. The solid line is the locus (equation (\ref{eqn:thinIPSC})) 
below which $\rho$ must fall if the immediate post-shock cooling radiation 
is to be negligible (as we assume).}
\end{figure*}

\subsection{Survival criterion}

We can now follow the evolution of a collapsing fragment. The opacity limit for the 
proto-fragment is reached as soon as $\mathcal{H}+\mathcal{D} \ga \mathcal{C}(T)$. 
In reality, the radiative cooling is more complicated than equation (\ref{eqn:cool}). The gas 
flowing into the accretion shocks which define the top and bottom of the fragment 
(see Fig. 1) is initially heated to high temperature 
($T_{_{\rm SHOCKED}} \sim 3\bar{m}v^2/16k_{_{\rm B}}$) in the shock front, and 
then rapidly cools behind the shock front. Provided we impose the condition 
$\rho v^3 / 2 \ll \sigma_{_{\rm SB}} T_{_{\rm SHOCKED}}^4\,$, or
\begin{equation} \label{eqn:thinIPSC}
\rho \;\ll\; \rho_{_{\rm CRIT}} \;\equiv\; \frac{3^3\,\pi^5\,\bar{m}^4\,v^5}
{2^{14}\,5\,c^2\,h^3} \;\simeq\; 10^{-12}\,{\rm g}\,{\rm cm}^{-3}\,\left[ 
\frac{v}{{\rm km}\,{\rm s}^{-1}} \right]^5 \,,
\end{equation}
then the immediate post-shock cooling (IPSC) radiation is either optically 
thin continuum radiation (e.g. from dust), or optically thick cooling in a few 
specific molecular lines which occupy a small total bandwidth. In either case, 
half the IPCS radiation is radiated away from the fragment, and half towards 
it. In the most extreme case, all of this latter half is absorbed by the 
fragment, and then has to be re-radiated at temperature $T$, along with the 
internal energy delivered by $PdV$ compression. Thus, by requiring the fragment 
to radiate at temperature $T$ {\it all} the energy dissipated in the accretion 
shock, we are making a rather conservative assumption.

To ensure that a proto-fragment forms a distinct condensation, i.e. that it 
continues to undergo contraction after the opacity limit has been reached, we 
set the following two survival conditions at the opacity limit:
\begin{enumerate}
\item{the proto-fragment must be contracting in both the radial and vertical 
dimensions, i.e. 
\begin{eqnarray} \label{eqn:survival1}
\dot{r}_{_{\rm LIMIT}} & < & 0 \,, \\ \label{eqn:survival2}
\dot{z}_{_{\rm LIMIT}} & < & 0 \,; 
\end{eqnarray}}
\item{the final extent  of the proto-fragment, in both dimensions, must be less 
than or equal to half its initial radial size, i.e.
\begin{eqnarray} \label{eqn:survival3}
r_{_{\rm LIMIT}} & < & 0.5\,r_{_{\rm INIT}} \,, \\ \label{eqn:survival4}
z_{_{\rm LIMIT}} & < & 0.5\,r_{_{\rm INIT}} \,.
\end{eqnarray}}
\end{enumerate}

For given $\rho$, $v$ and $a$ (i.e. for a given colliding flow), we must pick 
$t_{_{\rm INIT}}$ and $r_{_{\rm INIT}}$, where $t_{_{\rm INIT}}$ is the time 
at which the fragment under consideration starts to condense out, and 
$r_{_{\rm INIT}}$ is its initial radius. Thus the initial conditions for a 
trial fragment are
\begin{eqnarray}
t & = & t_{_{\rm INIT}} \,, \\
r(t_{_{\rm INIT}}) & = & r_{_{\rm INIT}} \,, \\
\dot{r}(t_{_{\rm INIT}}) & = & 0 \,, \\
z(t_{_{\rm INIT}}) & = & Z(t_{_{\rm INIT}}) \,, \\
\dot{z}(t_{_{\rm INIT}}) & = & \dot{Z}(t_{_{\rm INIT}}) \,.
\end{eqnarray}
We can then use equations (\ref{eqn:rddotnew}), (\ref{eqn:zddot}), (\ref{eqn:Pextz}), 
(\ref{eqn:Pextr}) and (\ref{eqn:mdot}) to follow the development of the fragment. 
Then we can use equations (\ref{eqn:heat}), (\ref{eqn:diss}) and (\ref{eqn:cool}) 
to evaluate ${\cal H}$, ${\cal D}$, ${\cal C}$, and check whether the fragment 
has reached the opacity limit (${\cal H} + {\cal D} \ga {\cal C}$). Finally, 
once the limit is reached, we can 
check the survival conditions, equations (\ref{eqn:survival1}), (\ref{eqn:survival2}), 
(\ref{eqn:survival3}) and (\ref{eqn:survival4}).

In order to reduce the parameter space we fix $a = 0.2\,{\rm km}\,{\rm s}^{-1}$, 
correponding to molecular gas at $10\,{\rm K}$. The equations are integrated 
using a fourth-order Runge-Kutta scheme.

In reality the gas in the two colliding streams will not have uniform 
density, and so $\rho$ should be interpreted as the mean pre-shock density. 
The presence of inhomogeneities in the pre-shock gas will promote fragmentation 
of the resulting 
layer, by creating the seed perturbations from which condensations 
subsequently grow. It will also tend to yield condensations with finite 
angular momentum. However, the lowest-mass condensations will be those 
with low angular momentum, and so we ignore angular momentum in this 
exploratory analysis.

\begin{figure*}
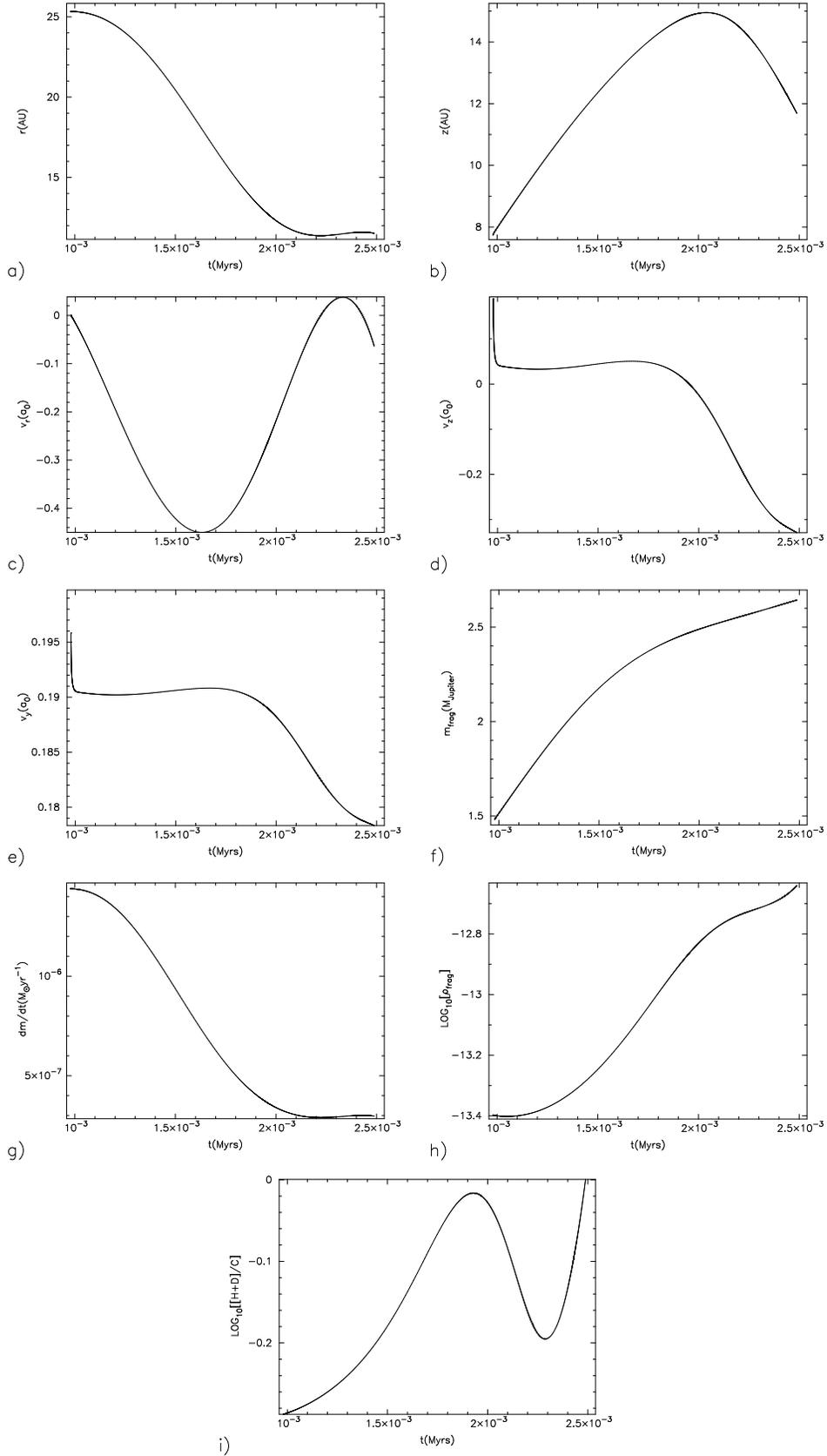
 
\label{fig:minmass}
\centering
 \includegraphics[angle=-90,width=6cm]{1703fg31.eps}
\vspace{0.1cm}
\hspace{0.3cm}
 \includegraphics[angle=-90,width=6cm]{1703fg32.eps}
\vspace{0.1cm}
\hspace{0.3cm}
 \includegraphics[angle=-90,width=6cm]{1703fg33.eps}
\vspace{0.1cm}
\hspace{0.3cm}
 \includegraphics[angle=-90,width=6cm]{1703fg34.eps}
\vspace{0.1cm}
\hspace{0.3cm}
 \includegraphics[angle=-90,width=6cm]{1703fg35.eps}
\vspace{0.1cm}
\hspace{0.3cm}
 \includegraphics[angle=-90,width=6cm]{1703fg36.eps}
\vspace{0.1cm}
\hspace{0.3cm}
 \includegraphics[angle=-90,width=6cm]{1703fg37.eps}
\vspace{0.1cm}
\hspace{0.3cm}
 \includegraphics[angle=-90,width=6cm]{1703fg38.eps}
\vspace{0.1cm}
\hspace{0.3cm}
 \includegraphics[angle=-90,width=6cm]{1703fg39.eps}
\vspace{0.1cm}
\caption{Evolution of the minimum-mass fragment: (a) radius, $r(t)$ in AU; (b) height, 
$z(t)$ in AU; (c) radial velocity, $\dot{r}(t)$ in units of $a = 0.2\,{\rm km}\,{\rm s}^{-1}$; 
(d) vertical velocity of material already in the fragment, $\dot{z}(t)$ in units of $a$; (e) 
vertical velocity of shock front bounding the fragment, $\dot{y}(t)$ in units of $a$; (f) 
mass of fragment, $m(t)$ in $M_{_{\rm JUPITER}}$; (g) accretion rate, $dm/dt$ in $M_\odot\,
{\rm yr}^{-1}$; (h) logarithmic fragment density, $\ell{\rm og}(\rho_{_{\rm FRAG}})$ in 
${\rm g}\,{\rm cm}^{-3}$; (i) logarithmic ratio of heating to cooling, $\ell{\rm og}(
[{\cal H}+{\cal D}]/{\cal C})$.} 
\end{figure*}

\section{Results and discussion}\label{sec:results}

\subsection{General considerations}

A condensing fragment posseses two orthogonal dimensions $r$ and $z$. Moreover, 
(a) $r$ and $z$ in general have different initial values ($r(t_{_{\rm INIT}})$ 
and $z(t_{_{\rm INIT}})$).  (b) Their excursions are driven by different external 
pressures (see equations (\ref{eqn:Pextz}) and (\ref{eqn:Pextr})). (c) The coupling 
between them (which occurs through both internal pressure and self-gravity) is 
non-linear. Therefore a fragment can evolve in quite complicated ways. In 
particular, small fragments which are not initially very unstable tend to 
oscillate until they accrete sufficient mass from the continuing inflow to 
become unstable against monotonic condensation. The oscillations in $r$ and 
$z$ normally have unrelated periods, and so the model has to be able to treat 
both oblate and prolate spheroids. However, once a fragment starts to condense 
out it is normally oblate, due to the extra ram pressure exerted in the $z$ 
dimension by the inflow.

\subsection{Oscillations} \label{sec:oscillations}

If we fix $\rho$ and $v$, and vary $t_{_{\rm INIT}}$ and $r_{_{\rm INIT}}$, 
we find that the lowest-mass condensation, $m_{_{\rm MIN}}(\rho,v)$, forms 
when $t_{_{\rm INIT}} \simeq t_{_{\rm FRAG}}$ and $r_{_{\rm INIT}} \simeq 
0.5\,r_{_{\rm FRAG}}$. Fragments with $r_{_{\rm INIT}} \la 0.5\,r_{_{\rm FRAG}}$ 
tend to undergo radial oscillations before accumulating sufficient mass to 
condense out. Fragments with $r_{_{\rm INIT}} > 0.5\,r_{_{\rm FRAG}}$ tend to 
condense out monotonically. The lowest-mass fragments are those which start 
with just enough mass to condense out monotonically, i.e. $r_{_{\rm INIT}} 
\simeq 0.5 r_{_{\rm FRAG}}$. Since $0.5\,r_{_{\rm FRAG}} \simeq R_{_{\rm JEANS2}}$, 
this finding agrees with the predictions of the linear stability analysis in 
\S\ref{sec:linfrag}.

\subsection{The optimum combination of $\rho$ and $v$ for producing low-mass 
condensations} \label{sec:optimum}

Suppose now that only $\rho$ is fixed, and $v$ is varied. For each $v$ we can 
determine the lowest-mass condensation $m_{_{\rm MIN}}(\rho,v)$, by varying 
$t_{_{\rm INIT}}$ and $r_{_{\rm INIT}}$ as described above. What we find is 
that as $v$ is increased at fixed $\rho$, $m_{_{\rm MIN}}(\rho,v)$ at first 
decreases, reaches a minimum, and then increases. Thus for any value of $\rho$ 
there is an optimum value of $v$ for spawning low-mass condensations. 
Fig. 2 shows the $(\rho,v)$ plane and marks discrete points 
for which $m_{_{\rm MIN}}(\rho,v)$ is less than $0.005 M_\odot \equiv 5 
M_{_{\rm JUPITER}}$. We see that there is a significant region of $(\rho,v)$ 
space in which condensations with mass below $0.005 M_\odot$ can form.

This region has an irregular boundary because proto-fragments can undergo 
out-of-phase pulsations in the $r$- and $z$-dimensions. If a proto-fragment 
undergoes a pulsation in the $r$-dimension and expands for a while ($\dot{r} 
> 0$), it experiences a prolonged period of growth due to accretion onto its 
relatively large cross-section ($\pi r^2$), and hence its mass increases 
rapidly. Since the onset, amplitude and phase of such pulsations is critically 
dependent on initial conditions, small changes in $\rho$ and/or $v$ can result 
in quite large changes in $m_{_{\rm MIN}}$.

The full line on Fig. 2 shows the locus $\rho = \rho_{_{\rm CRIT}} 
= 10^{-12}\,{\rm g}\,{\rm cm}^{-3}\,[v/{\rm km}\,{\rm s}^{-1}]^5\,$. Hence the 
condition for the IPSC radiation to be neglected (equation (\ref{eqn:thinIPSC})) 
requires $\rho$ to be well below this line. Evidently this condition is always 
easily fulfilled.

\subsection{The minimum mass}

The lowest-mass condensation of all is formed when $\rho \simeq 9.6 \times 10^{-16}\,
{\rm g}\,{\rm cm}^{-3}$, $v \simeq 5.1\,a \simeq 1.02\,{\rm km}\,{\rm s}^{-1}$, 
$t_{_{\rm INIT}} = t_{_{\rm FRAG}} \simeq 1000\,{\rm yr}$, and $r_{_{\rm INIT}} = 0.62\,
r_{_{\rm FRAG}} \simeq 25\,{\rm AU}$; the initial half-height of the fragment is 
$z(t_{_{\rm INIT}}) \simeq 8\,{\rm AU}$, and the initial mass of the fragment is 
$m(t_{_{\rm INIT}}) \simeq 1.5 M_{_{\rm JUPITER}}$. By the time the fragment has reached 
the opacity limit, its radius is $\sim 12\,{\rm AU}$, and it is roughly spherical, but 
slightly oblate. $1600\,{\rm yr}$ have passed since its inception, and its mass has 
increased by $1.1 M_{_{\rm JUPITER}}$ to $2.6 M_{_{\rm JUPITER}}$. Thus 
$M_{_{\rm MIN2}} \simeq 2.6\,M_{_{\rm JUPITER}}$.

Fig. 3 shows the development of this fragment. For the first $\sim 
1200\,{\rm yr}$, the fragment contracts in the radial dimension, but expands in the 
vertical dimension, due to accretion of extra material from the inflow into the layer. 
Around $1200\,{\rm years}$ after its inception the fragment starts to contract in the 
vertical dimension. At the same time there is a very small bounce in the radial dimension, 
but this is short-lived, and the fragment quickly resumes its radial contraction. 
Soon after this the opacity limit is reached, and we must presume that the fragment 
then switches to Kelvin-Helmholtz contraction. At this stage the fragment is roughly 
spherical. It is still accreting, at a rate $\sim 2 \times 10^{-4} M_{_{\rm JUPITER}}\,
{\rm yr}^{-1}$, and it has a dynamical timescale of $\sim 1000\,{\rm yr}$, so we 
might expect its mass to increase by a further few tenths of a Jupiter mass during 
the subsequent condensation.

\section{Conclusions}\label{sec:conclusions}

In 3-dimensional hierarchical fragmentation, the minimum mass is estimated to be 
$M_{_{\rm MIN3}} \sim 7\,{\rm to}\,10\,M_{_{\rm JUPITER}}$ (e.g. $7\,M_{_{\rm 
JUPITER}}$, Low 
\& Lynden-Bell 1976, $10\,M_{_{\rm JUPITER}}$, Silk 1977; $10\,M_{_{\rm JUPITER}}$, 
Boss 1988). Moreover, these estimates probably need to be revised upwards, to take 
account of ongoing accretion and merging during concensation (see \S\ref{sec:3D}). 

However, surveys of young clusters are beginning to find objects with 
masses estimated to be as low as $3\,M_{_{\rm JUPITER}}$ (e.g. Zapatero Osorio 
\emph{et al.} 2002). 

We find that, in the 2-dimensional fragmentation of a shock compressed layer, the 
minimum mass is significantly smaller than $10\,M_{_{\rm JUPITER}}$. For certain 
rather specific shock parameters, 
fragments with mass $< 3\,M_{_{\rm JUPITER}}$ can condense out -- specifically 
$M_{_{\rm MIN2}} \simeq 2.6\,M_{_{\rm JUPITER}}$ -- and for a wide 
range of shock parameters fragments with $\la 5 M_{_{\rm JUPITER}}$ can condense 
out. Since (i) layer fragmentation is probably a more realistic model for fragmentation 
in highly turbulent, transient star-forming clouds, (ii) layer fragmentation 
is more likely to be permanent than 3-dimensional fragmentation (because the 
fragments in a layer condense out faster than the layer as a whole), and (iii) 
our analysis takes proper account of ongoing accretion during fragmentation, we 
conclude that free-floating planetary-mass objects can form by the fragmentation 
of a shock-compressed layer, in the same way as stars and brown dwarves. 

\begin{acknowledgements}
DFAB thanks PPARC for the support of a postgraduate studentship. APW thanks the 
European Commission for support under contract HPRN-CT-2000-00155.
\end{acknowledgements}



\end{document}